# A peridynamic-informed deep learning model for brittle damage prediction

R. Eghbalpoor[1], A. Sheidaei[1,*]

[1]Aerospace Engineering Department, Iowa State University, Ames, IA, US

* Corresponding author, Email address: sheidaei@iastate.edu

## Abstract

In this study, a novel approach that combines the principles of peridynamic (PD) theory with PINN is presented to predict quasi-static damage and crack propagation in brittle materials. To achieve high prediction accuracy and convergence rate, the linearized PD governing equation is enforced in the PINN's residual-based loss function. The proposed PD-INN is able to learn and capture intricate displacement patterns associated with different geometrical parameters, such as pre-crack position and length. Several enhancements like cyclical annealing schedule and deformation gradient aware optimization technique are proposed to ensure the model would not get stuck in its trivial solution. The model's performance assessment is conducted by monitoring the behavior of loss function throughout the training process. The PD-INN predictions are also validated through several benchmark cases with the results obtained from high-fidelity techniques such as PD direct numerical method and Extended-Finite Element Method. Our results show the ability of the nonlocal PD-INN to predict damage and crack propagation accurately and efficiently.

## 1. Introduction

High-fidelity numerical techniques such as Continuum Damage Mechanics (CDM), Extended Finite Element Method (XFEM), Phase-Field Method (PFM), and Cohesive Zone Model (CZM) [1-6] have long been a mainstay in the study of fracture mechanics, where intricate physical events are precisely studied to understand crack initiation and propagation. Later, meshless methods were introduced to eliminate mesh dependency and re-meshing strategies. Within these methods, Peridynamic (PD) theory has emerged as a prominent alternative to Classical Continuum Mechanics (CCM), specifically tailored to handle material discontinuities such as cracks or heterogeneities across micro- to macro-structures [7-10]. Notably, PD has found successful applications in various domains, including brittle materials with elastic deformation [11, 12], rate-dependent inelastic deformations [13, 14], as well as heterogeneous and composite materials [15, 16]. It has proven effective in tackling challenges related to fatigue and impact loadings [17, 18], as well as multiphysics analysis [19, 20].

Despite the advantages of these high-fidelity techniques, computational demand arises with increasing structural and material complexity, especially when conducting extensive parametric studies or uncertainty quantification. To address these challenges, surrogate and machine learning (ML) models trained on high-fidelity simulations or experimental datasets, have emerged as effective solutions to maintain a balance between computational accuracy and efficiency [21-24]. Data-driven ML models can extract patterns and relationships directly from substantial datasets; however, their performance hinges on the quality and quantity of training data. Their susceptibility to overfitting the training data can hinder their effectiveness in extrapolating or handling scenarios beyond the training distribution.

Most recently, a new deep learning model known as Physics-Informed Neural Network (PINN) has emerged as a robust, low-to-no data-dependent ML tool to find a solution for various physical phenomena, including elasticity, fluid flow, heat, or sound propagation which are described and understood through solving the corresponding governing equations [25-30]. In PINN, governing equations, mainly Partial Differential Equations (PDEs), can be seamlessly integrated into the ML model loss function. This integration transforms the problem into an optimization task, effectively addressing the original physical problem through computational means. In the following, several studies including the application of PD in PINN will be reviewed. Niu et al. [31] enforced the governing equation of CCM for static equilibrium into the PINN's loss function consisting of individual losses for PDE, boundary conditions, and other constitutive relations to obtain finite-strain plasticity with isotropic hardening. Kamali et al. [32] utilized PINN to estimate the two-dimensional distribution of elastic modulus and Poisson's ratio from domain strain and normal stress boundary conditions. They validated the model with elasticity imaging experiments and numerical data from linear elastic materials under uniaxial loading. Haghighat et al. [33] introduced a PINN using the peridynamic differential operator (PDDO) as an alternative approach to the PD equations derivation in their nonlocal forms [34]. Their proposed PDDO-PINN method accurately captured elasto-plastic deformation of a square domain subjected to indentation by a rigid body. Ning et al. [35] employed PINN with a PD approach to predict the elastic displacement of homogenous and heterogeneous plates. They showed that utilizing the gradual-refinement sampling technique instead of direct fine sampling leads to enhanced accuracy and time efficiency in the model performance.

While data-driven deep NNs have previously been employed to capture damage and crack propagation [36-41], research focusing on PINN models in this context is limited. Dourado et al. [42] proposed a repeating recurrent NN to predict damage accumulation from corrosion-fatigue loading by incorporating a modified version of the Paris law, the Walker model, into the loss function to fine-tune network trainable parameters. Goswami et al. [43, 44] predicted crack propagation using phase-field theory integrated into the loss function of PINN. They achieved high accuracy with low computational cost by utilizing the transfer learning technique and re-training only the output layer of the network. Tu et al. [45] introduced the PointNet-based adaptive mesh refinement method to predict crack path in 2D plates with pre-crack by minimizing the variational energy of the system through the PINN optimization process. Zheng et al. [46] applied the principles of irreversible thermodynamics in CDM to the PINN loss function, where elastic residual energy of the damaged domain was minimized. They showed better convergence and robustness by decomposing the main domain into several subdomains and assigning separate networks to each subdomain. It is noteworthy to mention that despite the increasing utilization of PINNs in this field of study, a noticeable gap exists when exploiting the advantages of meshless and non-local methods such as PD.

This study explores a novel integration of the nonlocal PD method in conjunction with PINN to predict damage initiation and propagation in brittle materials. The main challenges in employing this method are the low convergence speed of training, low accuracy in interfaces and discontinuities, and the inefficacy of the PD-INN model in domains with high-density Material Points (MPs). Therefore, the proposed PD-INN is rigorously assessed through diverse case studies encompassing a spectrum of MP densities — from low to high. A combination of the loss function with the linearized PD governing equation is proposed to increase accuracy and convergence speed. Since obtaining the PD forces introduces a considerable computing overhead, network architectures and hyperparameters are thoroughly assessed to reduce the required training epochs. Furthermore, deformation-gradient aware optimization technique is utilized while training high-density domains to prevent PD-INN from becoming trapped in its trivial solution. Eventually, the transfer learning technique is utilized to reduce the computational time needed for the subsequent time increments.

The remaining sections of this paper are structured as follows: In Section 2, an introduction to the theory of bond-based PD is provided, explaining the linearized form of its governing equation. This section also presents the PD-INN, covering a detailed explanation of the residual-based loss function and the details of the model hyperparameters. In section 3, detailed discussions results are provided and followed by comparisons between the predictions and true deformations. Finally, in section 4, concluding remarks are presented.

## 2. Methods and Implementation

In PD theory, the inherent non-local character of the method is a key feature. This arises from the fact that each material point (MP) is influenced by its neighboring points, incorporating length-scale parameters and long-range force interactions. Essentially, in PD, each MP interacts with points within a predefined radius, referred to as the horizon, forming a sphere (or circle in 2D). In this theory, a domain is discretized into particles with pairwise force interactions occurring through bonds inside the horizon. This discrete approach introduces an integral form of the equation of motion, distinguishing it from the differential equations found in CCM. This makes PD capable of dealing with problems including discontinuity and heterogeneity. By assuming small deformations and removing the time-derivative term in the equation of motion, linearized form of the PD governing equation can be implemented to find the quasi-static solution of the PD governing equation. The linearized PD governing equation is included in building the loss function in addition to other common residual-based loss terms such as boundary conditions and internal forces. In the subsequent subsections, a detailed description of the linearized PD as well as PD-informed NN is provided.

*2.1 Linearized bond-based peridynamic*

The equation of motion in PD is achieved by considering the force balance of a MP **x** [7, 8]:

$$(\rho dV_x)\ddot{\vec{\mathbf{u}}}(\mathbf{x}',t) = \int_{\beta_x}\left[\vec{\mathbf{f}}(\mathbf{q}',\mathbf{x}',t)dV_\mathbf{q}\right]dV_\mathbf{x} + \vec{\mathbf{b}}(\mathbf{x}',t)dV_x \tag{1}$$

where $\beta_x$ is the domain in the horizon of **x**, $\rho$ is density, $\mathbf{b}(\mathbf{x}',t)$ is body force density applied, $\vec{\mathbf{f}}(\mathbf{q}',\mathbf{x}',t)$ is pairwise bond force density exists in the horizon of **x** and is governed by constitutive equations, $dV_{\mathbf{x},\mathbf{q}}$ are volumes of the points in interaction, and $t$ is time. Symbol (') stands for the deformed configuration. The integral term of the equation is to maintain interactions of all MPs within the horizon and is substituted by a summation of all bond forces that exist in **x**'s neighborhood. MPs near the boundaries, where the horizon area is only partially defined, exhibit slightly softer mechanical properties compared to those within the interior of the domain. To address this issue, the commonly employed Geometry Modification method [47-49] is utilized. This method involves the addition of a layer of fictitious material points, typically with a thickness equal to the horizon, to the main domain where displacement boundary conditions are applied.

Bond-based PD is one of the most common and widely used material models in which two MPs exert the same amount of load in their bond direction [50-52]. The value of bond force is obtained as follows:

$$\vec{\mathbf{f}}(\mathbf{q},\mathbf{x}) = \omega\, c\, s\, (1-\mu)\vec{\mathbf{M}},\ c = \frac{9E}{\pi t \delta^3},\ s = \frac{e}{\xi},\ \omega = e^{-4\xi^2/\delta^2},\ \mu = \begin{cases} 1 & s \geq s_{cr} \\ 0 & \text{otherwise} \end{cases} \quad (2)$$

where $c$ is bond stiffness (or micromodules), $E$ is Young's Modulus, $t$ is thickness and $\delta$ is horizon radius equal to 3 times grid space. The presented equation for micromodules is for 2D plane stress. 2D plane strain or 3D constitutive models of $c$ can also be found in the literature. $s$ is the bond stretch which is obtained by the division of bond elongation $e$ to the initial length of the bond, $\xi$. $\omega$ is an influence function or length correction factor to reduce the interaction effects of MPs when $\xi$ increases. $\mu$ is brittle damage parameter which is equal to 1 when the stretch is beyond the critical stretch $s_{cr}$ and 0 otherwise. In this study, $s_{cr}$ is obtained by known fracture energy $G_0$ as $s_{cr} = \sqrt{4\pi G_0/9E\delta}$. Finally, $\vec{\mathbf{M}}$ is unit vector of bond direction.

The analytical solution of the PD integro-differential motion equation (1) is not feasible, necessitating the utilization of numerical techniques for both time and space discretization. While explicit time integration schemes with controlled time steps are required for dynamic loadings [10, 53], quasi-static analyses are treated with different approach, mainly iterative or direct methods where dynamic effects like wave propagation are diminished and larger time steps are used.

Notably, Adaptive Dynamic Relaxation method (ADR) was introduced by Kilic and Madenci [54] to obtain steady-state solution of the dynamic PD equation, knowing the fact that the transient response will converge to its steady-state condition which is equal to static solution of the problem. In this method, an artificial damping term is added to the governing equation of PD, Eq. (1), and central-difference iterative method is utilized to find the steady-state solution after a number of iterations [11, 12]. However, this approach can be time-consuming when dealing with crack propagation problems since loads are applied incrementally and, in each increment, a certain number of iterations is required to converge to the steady-state solution. Moreover, identifying the most effective damping coefficient is not always straightforward, and improper numerical parameters may lead to a low convergence rate. To mitigate these challenges, direct solution is proposed by solving the system of equation $\mathbf{Ku}=0$ where $\mathbf{K}$ represents the global stiffness matrix and $\mathbf{u}$ denotes the displacement vector of all MPs [55, 56]. In order to find stiffness matrix, the linearized form of the governing equation with respect to the displacements $\mathbf{u}$ can be derived. This method will be reviewed here. According to Figure 1, elongation $e$ can be obtained as $\left|\vec{\xi}+\vec{\eta}\right|-\left|\vec{\xi}\right|$ where $\vec{\eta}=\vec{u}_q - \vec{u}_x$ is relative displacement. Therefore, $s = \left(\left|\vec{\xi}+\vec{\eta}\right|-\left|\vec{\xi}\right|\right)/\left|\vec{\xi}\right|$ and $\vec{M} = \left(\vec{\xi}+\vec{\eta}\right)/\left|\vec{\xi}+\vec{\eta}\right|$. The first step to linearize $\vec{f}(\mathbf{q'},\mathbf{x'})$ with respect to $\vec{\eta}$ is to write the first order Taylor's series expansion by considering a small-near-to-zero $\eta = \left|\vec{\eta}\right|$. Thus, we have:

$$\vec{f}(\mathbf{q'},\mathbf{x'}) = \vec{f}(\mathbf{q'},\mathbf{x'})\Big|_{\eta=0} + \frac{\partial \vec{f}(\mathbf{q'},\mathbf{x'})}{\partial \mathbf{\eta}}\Bigg|_{\eta=0} \eta \tag{3}$$

where $\vec{f}(\mathbf{q'},\mathbf{x'})\Big|_{\eta=0} = \vec{f}(\mathbf{q},\mathbf{x}) = 0$ states the undeformed configuration. Evaluating the second term of Taylor's expansion and substituting $\eta = 0$, Eq. (3) can be written as [57]:

$$\vec{f}(\mathbf{q'},\mathbf{x'}) = \frac{\partial \vec{f}(\mathbf{q'},\mathbf{x'})}{\partial \mathbf{\eta}}\Bigg|_{\eta=0} \eta = \omega\, c\, (1-\mu) \frac{\vec{\xi}\otimes\vec{\xi}}{\left|\vec{\xi}\right|^3} \mathbf{\eta} \tag{4}$$

where $\otimes$ is dyadic product. Eq. (4) can be shown in vector representation as:

$$\vec{f}(\mathbf{q}',\mathbf{x}') = \frac{\omega c (1-\mu)}{|\vec{\xi}|^3} \left( \begin{pmatrix} \xi_1 & \xi_2 \end{pmatrix} \cdot \begin{pmatrix} \eta_1 \\ \eta_2 \end{pmatrix} \right) \begin{pmatrix} \xi_1 \\ \xi_2 \end{pmatrix} \tag{5}$$

in this equation, (.) is vector dot product and subscripts 1,2 are directional components (in 2D). Final step is to rewrite the Eq. (5) in terms of the pairwise particle displacements $\vec{\mathbf{u}}_x$ and $\vec{\mathbf{u}}_q$:

$$\vec{f}(\mathbf{q}',\mathbf{x}') = \frac{\omega c (1-\mu)}{|\vec{\xi}|^3} \begin{pmatrix} \xi_1\xi_1 & -\xi_1\xi_1 & \xi_1\xi_2 & -\xi_1\xi_2 \\ \xi_2\xi_1 & -\xi_2\xi_1 & \xi_2\xi_2 & -\xi_2\xi_2 \end{pmatrix} \cdot \begin{pmatrix} u_{q,1} & u_{x,1} & u_{q,2} & u_{x,2} \end{pmatrix}^T \tag{6}$$

where $(\cdot)^T$ is vector transpose. By eliminating the acceleration term and substituting the integral term by the summation of all bond-forces inside the horizon of **x**, Eq. (1) is rewritten as $\sum_{\beta_x} \vec{f}(\mathbf{q}',\mathbf{x}') dV^2 = 0$. Eventually, stiffness matrix can be easily derived by using the Eq. (6). Please note that it is assumed the domain is discretized uniformly, so $dV_x = dV_q = dV$ and no body force is applied. A similar approach in FEM can be applied here to eliminate known boundary displacements from the system of equations and subsequently derive the right-hand-side (**rhs**) vector to solve the unknown deformations.

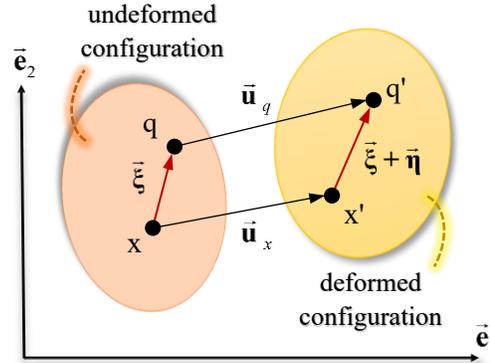

Figure 1. a schematic presentation of a PD bond, before and after a small deformation.

*2.2 Peridynamic-informed Neural Network*

In this section, the details of the proposed PD-informed NN are discussed. PINNs can be embodied with thousands-to-millions of parameters, enabling them to provide more accurate results while capturing nonlinear patterns through multiple transformations from the input layer to the output layers. Therefore, fully connected deep NN (FCNN) is used to build our proposed PD-INN and conduct feed forward process of the input variables, here the coordinates of all MPs, and train the network through backpropagation algorithm. Typically, PINNs consist of an input layer (L=0), $l$ hidden layers (L=1:$l$) and an output layer (L=$l$+1). Each layer contributes its data as input to the subsequent layer, having processed the output of each unit through an activation function which can be represented as:

$$z^L = \varphi\left(\mathbf{W}^{L-1}\mathbf{z}^{L-1} + \mathbf{b}^{L-1}\right), \quad L=1:l \tag{7}$$

where $z^L$ and $z^{L-1}$ are the output of ($L$)th and ($L$-1)th layer, respectively, $\mathbf{W}^{L-1}$ and $\mathbf{b}^{L-1}$ are the concatenated weights and biases of previous layer, $\varphi$ is activation function which is *tanh* for all layers except it is linear for the output layer. The network's output should satisfy the PD constraints in order to predict a precise deformation. Consequently, the following total loss function, denoted as $\mathcal{L}_{total}$, is constructed to provide FCNN with the knowledge of PD governing equation:

$$\mathcal{L}_{total} = \mathcal{L}_{PD\_forces} + \mathcal{L}_{B.C.s} + \mathcal{L}_{linearized} + \left(\mathcal{L}_{true\_data}\right) \tag{8}$$

where $\mathcal{L}_{PD\_forces}$ is the loss corresponding to the forces of internal material points and free boundaries, $\mathcal{L}_{B.C.s}$ is loss due to the applied boundary conditions like Dirichlet BCs, $\mathcal{L}_{linearized}$ is the loss for the residual of the linearized PD, and finally, $\mathcal{L}_{true\_data}$ is the (optional) data-driven loss according to the known data. A detail description of these losses is provided in the following equation:

$$\begin{aligned}
\mathcal{L}_{PD\_forces} &= \mathcal{L}_{int\_forces} + \mathcal{L}_{def.\ grad} + \mathcal{L}_{dmg} \\
&= \left\| \int_{\beta_x} \left[\vec{\mathbf{f}}(\mathbf{q}',\mathbf{x}',t)dV_q\right]dV_x \right\|, \quad \mathbf{x} \in \Omega_1, \Omega_2, \Omega_3 \\
\mathcal{L}_{B.C.s} &= \left\| \mathbf{u}_{BC\_pred.} - \mathbf{u}_{BC}^* \right\| \\
\mathcal{L}_{linearized} &= \left\| \mathbf{K}\mathbf{u}_{pred.} \right\| \text{ or } \left\| \mathbf{K}^{-1}.\mathbf{rhs} - \mathbf{u}_{pred.} \right\| \\
\mathcal{L}_{true\_data} &= \left\| \mathbf{u}_{pred.} - \mathbf{u}^* \right\|
\end{aligned} \tag{9}$$

where $\Omega_1$ is forces of all internal MPs except those with boundary conditions applied, $\Omega_2$ is the forces of MPs with higher deformation gradient and $\Omega_3$ correspond to damaged MPs. (*) shows the known displacements either from the applied boundary conditions or true numerical/experimental data. $\ell^2$-norm function $\|\bullet\|$ is implemented to calculate each loss term's value. Two options exist for constructing the $\mathcal{L}_{\text{linearized}}$: the first involves calculating $\|\mathbf{K}\mathbf{u}_{\text{pred.}}\|$ as the loss, which has the same order of magnitude as $\mathcal{L}_{\text{PD\_forces}}$. The second option is to compute the residual of $\|\mathbf{K}^{-1}.\mathbf{rhs} - \mathbf{u}_{\text{pred.}}\|$ which shares a similar order of magnitude with losses $\mathcal{L}_{\text{B.C.s}}$ and $\mathcal{L}_{\text{true\_data}}$. The choice made here is important to build the weighted some of all losses. Here, we followed the second choice which found to enhance model performance better. It should also be noted that, while $\mathcal{L}_{\text{true\_data}}$ is not a general requirement in PINNs, its presence would further accelerate model convergence rate. Once the weighted sum of all these losses is obtained, it is used to perform backpropagation algorithm. This is conducted using the automatic differentiation approach which calculates the gradient of the $\mathcal{L}_{\text{total}}$ with respect to the model trainable parameters $\mathbf{W}$ and $\mathbf{b}$. To this end, one of the most used and open-source ML packages, called TensorFlow [58], is used to calculate such gradients. This library supports GPU accelerated operations via Nvidia CUDA which can increase training speed from several to a hundred times in comparison to CPU implementation.

As stated earlier, the order of magnitude of the loss $\mathcal{L}_{\text{PD\_forces}}$ presented in Eq. (8) differs from that of either $\mathcal{L}_{\text{B.C.s}}$ or $\mathcal{L}_{\text{true\_data}}$ given their distinct units. Therefore, the loss minimization process is basically a multi-objective optimization process and is more likely to become trapped in the local minima of a non-convex loss function. To address this challenge, the balancing of different loss terms' contribution in $\mathcal{L}_{\text{total}}$ is required. Therefore, the weighted sum of the different terms is recommended for controlling the magnitude of each term. Additionally, an alternative approach involves monitoring and adjusting the gradient directions corresponding to each term, which can enhance both time efficiency and accuracy. However, it's worth noting that this adjustment becomes less critical when $\mathcal{L}_{\text{linearized}}$ or $\mathcal{L}_{\text{true\_data}}$ are included.

Finally, an extension of stochastic gradient descent (STG) known as the Adam optimizer is employed to perform fine-tuning of the model trainable parameters. The Adam optimizer dynamically adjusts the learning rate for each trainable parameter based on the history of gradients.

This adaptive approach aids in achieving faster convergence and improved accuracy compared to using a fixed learning rate. Additionally, it is beneficial to reassign and reduce the initial learning rate throughout the training procedure by implementing a schedule decay rate. This strategy allows the algorithm to descend towards the global minimum, mitigating the risk of oscillations around this critical point.

Figure 2 provides a representation of the steps undertaken by the PD-INN to predict the deformation within an arbitrary domain subjected to loading conditions. The initial step is to discretize the domain into a series of material points, where their coordinates besides other geometrical parameters like pre-crack length and position (if applicable) are served as inputs to the first hidden layer. Subsequently, the deformation components of all material points in the domain are generated as outputs, which in turn are employed to calculate the overall loss. Identifying the MPs to which boundary conditions are applied, those MPs located within the domain, global stiffness matrix **[K],** and the neighbors inside the horizon of each MP would make it possible to compute $\mathcal{L}_{\text{total}}$. Eventually, once the value of total loss reaches a threshold $\mathcal{L}_{\text{thr}}$, the training is stopped, and the final deformation is obtained.

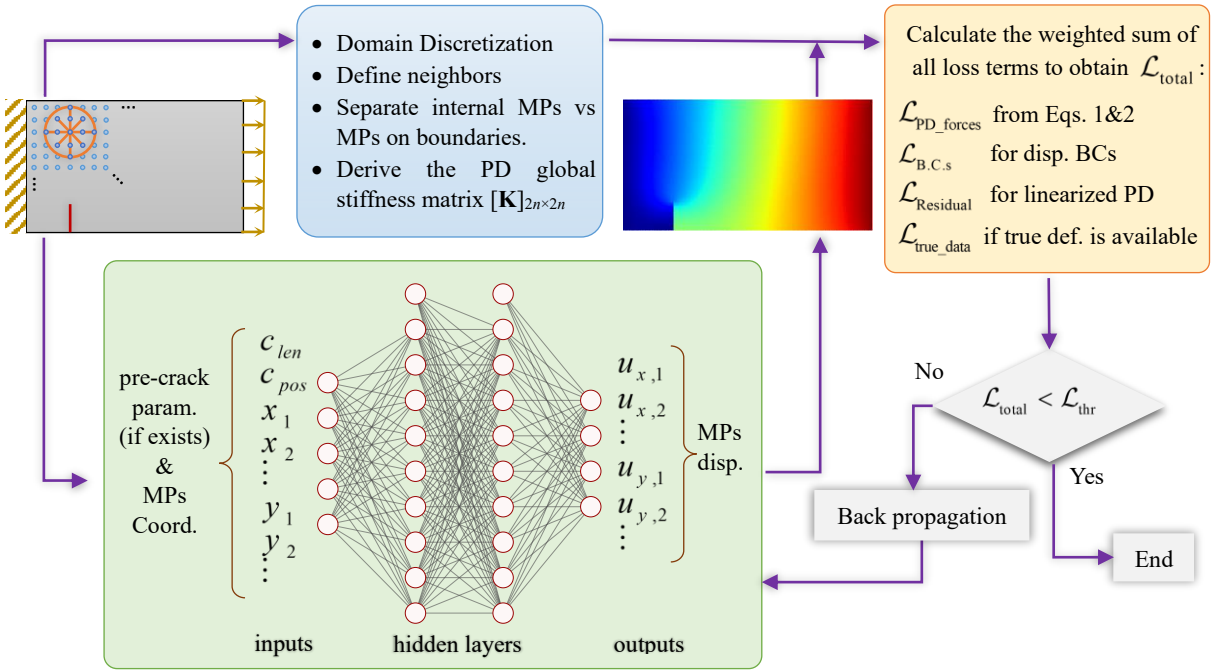

Figure 2. Schematic representation of the PD-INN model training process in this study

## 3. Results and Discussion

Here, the results obtained from PD-INN are validated through several benchmark problems and compared with the ground truth deformation from high-fidelity techniques such as PD direct numerical method, FEM and X-FEM. In section 3.1, several network architectures are investigated to identify those that offer superior accuracy and a fast convergence rate. In section 3.2, the PD-INN is trained to predict the deformation of a 2D plate with various pre-crack lengths and positions. The training loss curve is also presented to assess the effect of proposed loss function and hyperparameters on the model's accuracy and convergence rate. In section 3.3, the PD-INN is further evaluated to predict deformation of a domain consisting of stress localization around a circular cutout.

*3.1 PD-INN architecture*

The objective of this analysis was to discover the network configuration that would meet the desired accuracy, $\mathcal{L}_{thr}$, in the most time efficient way. Therefore, the PD-INN was trained to predict the deformation of a 2D plate with an arbitrary pre-crack as depicted in Figure 4. This analysis encompassed the choices over the network's input/output shape, number of hidden layers, and the units within each layer. Table 1 shows the details and specification of each network architecture, including the number of training parameters and the runtime corresponding to each network. DOF is the degree of freedom, equal to 2 for 2D analyses, and nMP is the number of material points within the domain.

Table 1. Various networks of choice to build PD-INN and their statistics

| Network | Input/Output shape | # Hidden layers | # units | # trainable parameters | $\mathcal{L}_{total} < \mathcal{L}_{thr}$ | Runtime* |
|---|---|---|---|---|---|---|
| i | DOF | 6 | 60 | 18602 | No | > 6 hr |
| ii | DOF | 6 | 100 | 51002 | No | > 6 hr |
| iii | DOF | 10 | 60 | 33242 | No | > 6 hr |
| iv | DOF | 10 | 100 | 91402 | No | > 6 hr |
| v | DOF | 15 | 150 | 317852 | No | > 6 hr |
| vi | DOF×nMP | 6 | 60 | 1240460 | Yes | ~ 2 hr, 50 min |
| vii | DOF×nMP | 6 | 100 | 2080700 | Yes | ~ 2 hr, 55 min |
| viii** | DOF×nMP | 6 | 60 | 1240460 | Yes | ~ 15 min |

In all architectural configurations detailed in Table 1, excluding network viii, the Glorot (Xavier) normal initializer and the hyperbolic tangent (*tanh*) activation function were adopted for constructing the FCNN. To fine-tune the network's trainable parameters, Adam optimizer was employed with an initial learning rate set at 5e-4 and a schedule decay rate 0.9 applied every 1000 epochs. The optimal learning rate for the networks i-v was observed to fall within the range of 5e-4 to 5e-3, although they failed to converge to the target threshold loss; values exceeding or falling below this range resulted in decreased network performance. In contrast, an examination of the training loss curve, as illustrated in Figure 3, reveals that this learning rate range would make the networks vi and vii struggle to minimize the loss and lead to considerable fluctuations during the training. For consistency in this comparative analysis, identical configurations are maintained. It is evident, both from Table 1 and Figure 3, that a reduction in the required number of epochs to attain the threshold loss has a noticeable effect on the overall computational time. This phenomenon can be attributed to the relatively more time-consuming nature of PD internal force calculations as opposed to gradient computations and network parameter updates. Observing almost similar performance in networks vi and vii, network vi architecture is selected to conduct analyses in the next subsections. Detailed information on the network viii performance, which exhibits a significant enhancement in convergence rate and training speed, is provided in section 3.2.

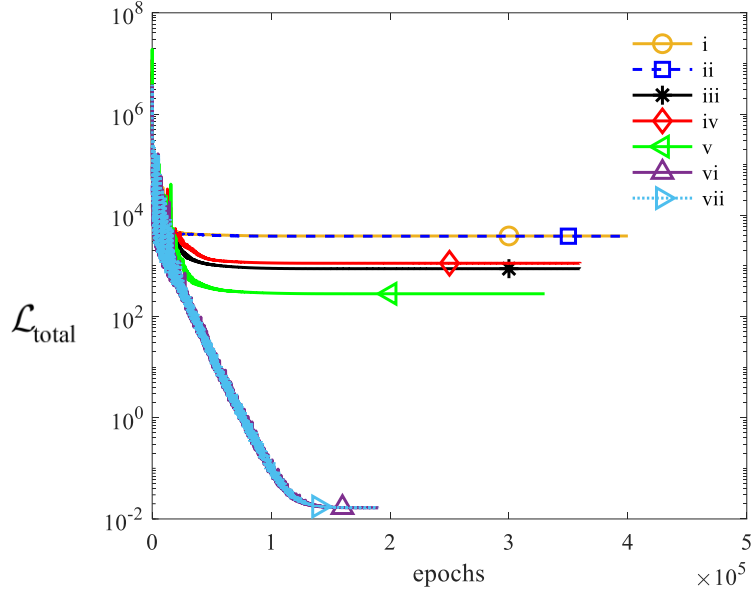

Figure 3. Various PD-INN architectures and their performance on a plate with an arbitrary discontinuity

## 3.2 Plates with pre-crack

As it was determined in section 3.1, network vii architecture with input/output shape of DOF×nMP, and six hidden layers with 64 units in each layer is applied to build the FCNN. A training dataset consisting of the true deformations of a rectangular plate (W×H) 0.1m×0.05m under tension was used to train the network with hybrid loss function. The training dataset includes ground truth deformation associated with six different vertical pre-cracks of length 10mm and 30mm at positions x=0.02m, 0.05m and 0.08m. It should be noted that given the limited amount of training data in this study, the objective is not to create a surrogate model. Furthermore, a more complex ML model like parallel FCNNs or convolutional neural network (CNN) is required to build such a model. The purpose of this training step is to accelerate model training and create a sensitivity concerning the geometrical parameters of crack lengths ($c_{len}$) and positions ($c_{pos}$) as shown in Figure 4. Thus, these parameters are combined with the input layer of the FCNN. Once the network is trained, it is fine-tuned to meet the threshold training loss and predict an accurate deformation. Figure 4 presents the displacement contours with the mean absolute error (MAE) and the error contour of the relative final deformed configuration for three different cases. Each domain is under uniaxial tension of U1=0.2mm where it is ensured no further damage will happen. The following equation calculates the MAE:

$$\text{MAE} = \frac{\sum_{i=1}^{n} \left| U_{\text{Predication}}^{i} - U_{\text{true\_data}}^{i} \right|}{n} \qquad (10)$$

where $U$ is displacement and $n$ is the total number of MPs. According to the results presented in Figure 4, the PD-INN prediction is in good agreement with true deformation where the MAE is as low as three orders of applied displacement boundary condition.

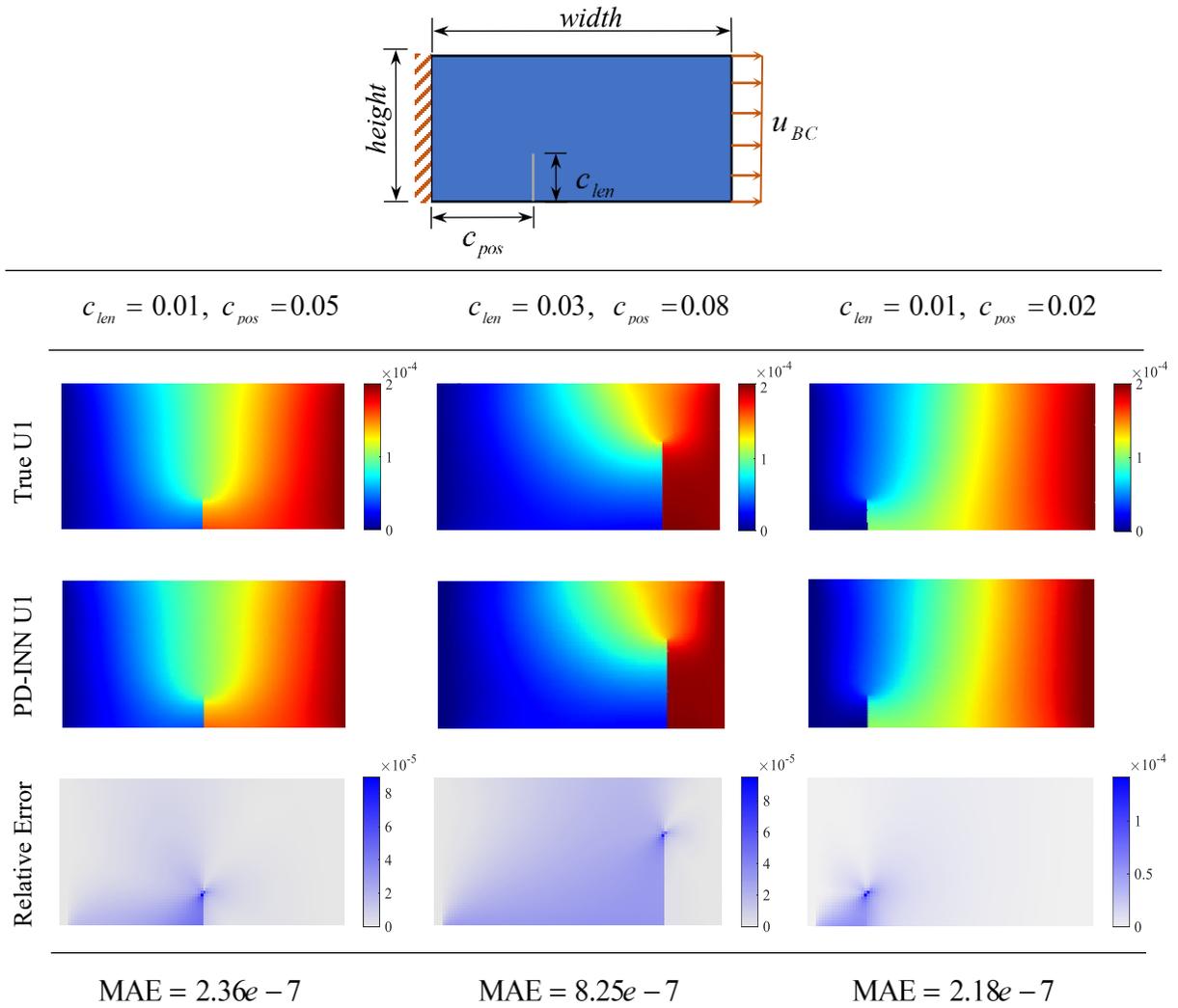

Figure 4. PD-INN prediction of deformation U1 in presence of pre-crack b) ground truth U1 obtained from PD direct method implementation

Upon close evaluation of the relative error contour presented in Figure 4, the area on the left side of the pre-cracks exhibits higher error in comparison to its other side. This observation stems from the propagation hypothesis [59, 60] which states that in PINNs, the boundary conditions propagate from boundaries to the inner domain. It is also observed that the optimizer may stop this propagation and get stuck at its trivial solution $u = 0$ for the rest of the domain. As a result, a narrow region of a high deformation gradient will exist near the boundaries. This issue is more probable in domains with higher MP densities. While building the hybrid loss function with $\mathcal{L}_{\text{true\_data}}$ would mitigate this issue, further regularization and consideration should be made in case of an unseen case without ground truth. Low speed of convergency is another concern regarding the cases where $\mathcal{L}_{\text{true\_data}}$ is removed from the loss function. To address these issues, after extensive evaluation of the network's performances, the following enhancements was found to be effective for the framework of the PD-INN:

- Random Normal kernel initializer with 0.001 standard deviation was selected over the common choice of Glorot (or Xavier) initializer. This choice was made to specifically minimize the scattering of initial domain predictions.
- Initial learning rate equal to 1e-5 with an exponential decay learning rate of 0.9 is considered every 5e2 epochs.
- Forcing regions with higher deformation gradients by considering the $\mathcal{L}_{\text{def. grad}}$ to ensures smooth propagation from boundaries to the inside of the domain. Common masks for computing the components of gradient in a 2×2 neighborhood [61] is $M_x$=[-1 1;-1 1] or $M_y$=[1 1;-1 -1]. In order to generalize this approach to various domains with or without crack, a cyclical annealing schedule [62] is found to be effective which considers the factor $\beta$ in an increase-reset-to-zero pattern every 5e3 epochs as depicted in Figure 5.
- The choice of $\|\mathbf{K}^{-1}.\mathbf{rhs} - \mathbf{u}_{\text{pred.}}\|$ for computing the $\mathcal{L}_{\text{linearized}}$ serves similar to $\mathcal{L}_{\text{true\_data}}$ in the training process which makes it a better choice than $\|\mathbf{K}\mathbf{u}_{\text{pred.}}\|$, as presented in Eq. (9). Therefore, adding this form of $\mathcal{L}_{\text{lineraized}}$ makes the network more stable and increases the convergency rate considerably.

Figure 5 presents the loss curve and its composition with different terms for a domain with crack length $c_{len} = 0.2$ and position $c_{pos} = 0.35$, respectively. By considering the model with all these proposed enhancements, a high convergence rate as well as accuracy was achieved. In order to better indicate the influence of the proposed linearized PD loss, the training is performed with ($\gamma=1$) and without ($\gamma=0$) $\mathcal{L}_{\text{linearized}}$. The results indicate that attaining the desired accuracy with $\gamma=0$ requires more than 50e3 epochs of training. In contrast, approximately 17e3 epochs were found to be sufficient when using $\gamma=1$, although the potential for achieving even higher accuracy remains. It can also be seen from the Figure 3 and Figure 5 that having a PD-INN sensitive to the pre-crack parameters resulted in the considerably reduced loss value at the beginning of the training, where the displacement propagation from boundaries has already happened.

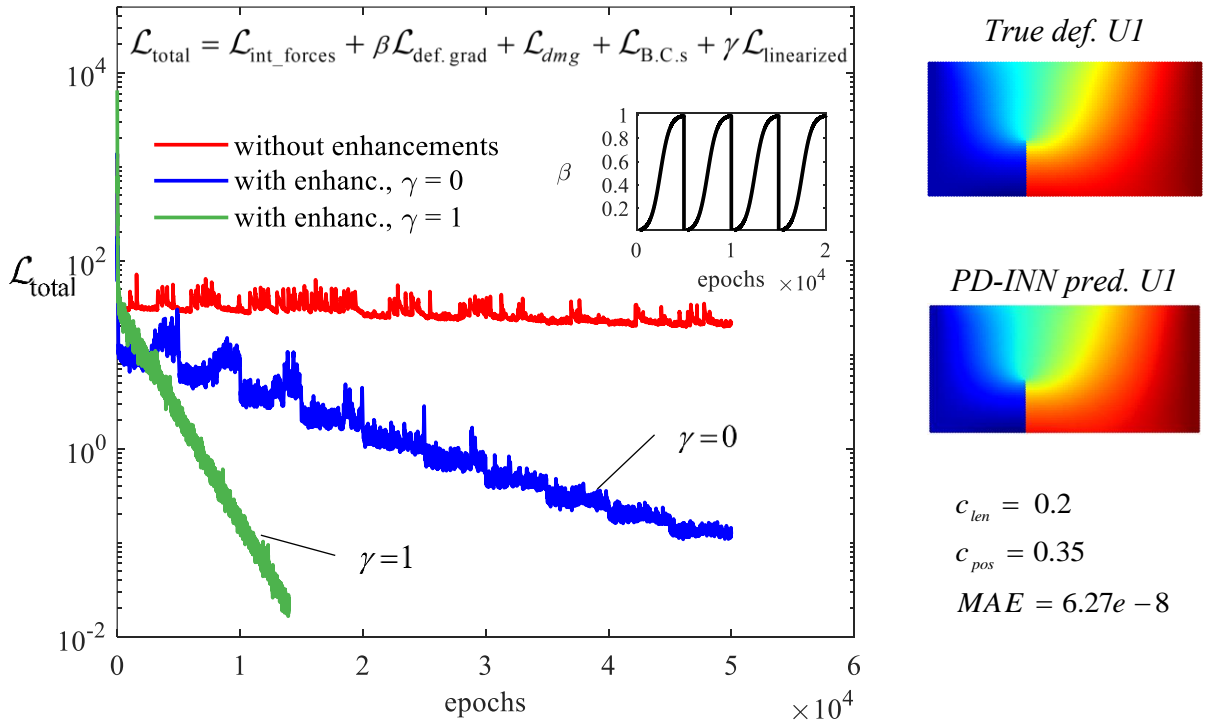

Figure 5. Comparison of training loss for training with or without proposed enhancements

Once the PD-INN is trained, it is prepared to be implemented for the damage analysis. The trained model parameters are transferred to a new network for each following time increment using the transfer learning technique. This would decrease the training efforts considerably compared to a network without having a previous understanding of the physics of the problem. In addition, it is found that modifying the network so that only the parameters assigned to the output layer are trainable would be sufficient in converging to the final deformation. Considering all these methods as well as previously suggested enhancements would lead to a fast and accurate prediction in crack propagation and evaluation of the strength of the material. The crack propagation using PD-INN for the case $c_{len} = 0.2$ and $c_{pos} = 0.35$, shown in Figure 5, is investigated and compared to the true crack path obtained from direct method of calculation. Therefore, the plate is subjected to the uniaxial tension with the rate of 0.02 mm.s$^{-1}$, material property $E$=192 GPa and fracture energy is $G_0 = 83 \, \text{kJ/m}^2$, grid space 1 mm and horizon radius of 3 mm. Figure 6 presents the PD-INN crack path prediction and Force-Displacement curve. The PD-INN predictions are in good agreement with the results obtained from PD direct solution, with almost 3% error in stiffness calculation. It's worth noting that in this problem, the crack has not grown vertically due to the off-center position of the pre-crack and, as a result, existence of bending moments around the crack tip.

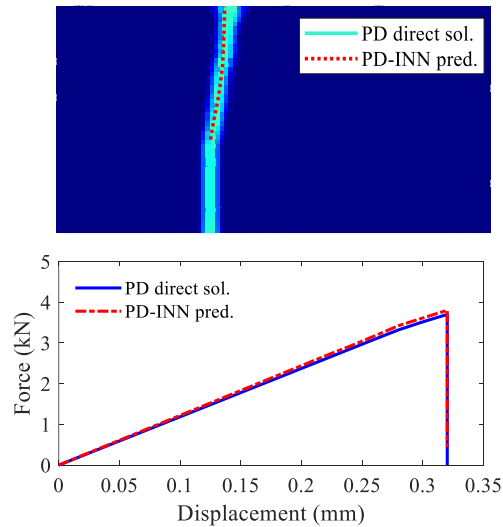

Figure 6. The comparison of crack path (top) and force-displacement curves (bottom) obtained from PD-INN and direct PD solution

Next, a more complex case study is conducted to assess the PD-INN's predictive capabilities concerning crack propagation within a square plate containing ten randomly located pre-cracks. Thus, a square plate of width 2 in, grid space 0.015 in (with approx. 18600 MPs) is modeled as presented in Figure 7(a). The plate is stretched vertically until a complete failure happens and the final crack path is compared with the result obtained from Extended-FEM (XFEM) method [63] as depicted in Figure 7(b). In both methods, crack propagation originated from the central pre-crack, eventually merging with two adjacent pre-cracks situated on either side. Subsequently, the crack propagated from the tips of these two pre-cracks and extended towards the domain's boundary. Overall, our findings demonstrate a good agreement between the outcomes obtained through both methods.

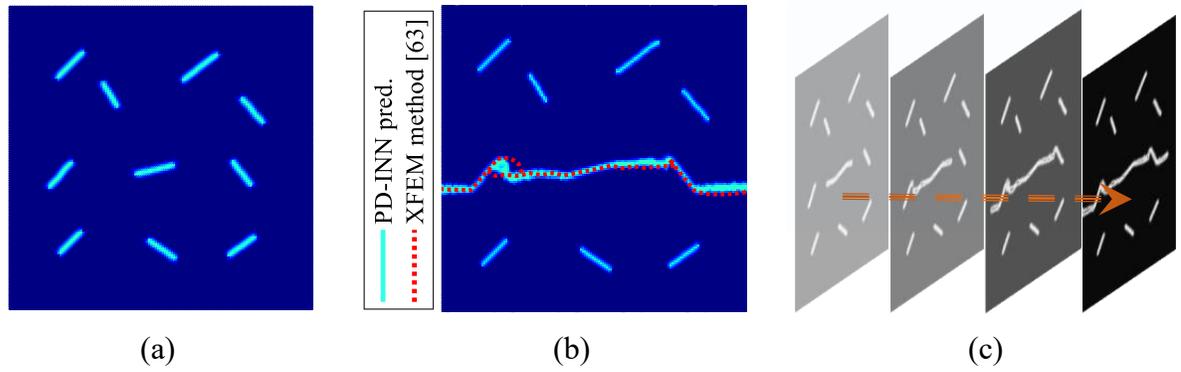

Figure 7. A comparison of crack path predicted by PD-INN and XFEM [63] in a square plate with random pre-cracks (a) Initial configuration (b) Final crack path (c) sequences of crack growth

*3.3 A plate with circular cut-out*

To further evaluate the effectiveness of the proposed PD-INN model in a domain with no initial cracks, a 50mm×50mm square plate with the thickness of 0.5mm containing a central D=20mm circular hole was considered. Horizon radius of 0.6 mm, fracture energy of $G_0 = 83\,\text{kJ/m}^2$, and critical stretch $s_{cr}$ = 3.16e-2 was implemented. As illustrated in Figure 8-(a), a quarter of the domain is modeled where symmetry boundary condition was applied to both sides. The domain was stretched until complete failure occurred, and the results were compared with PD direct numerical solution.

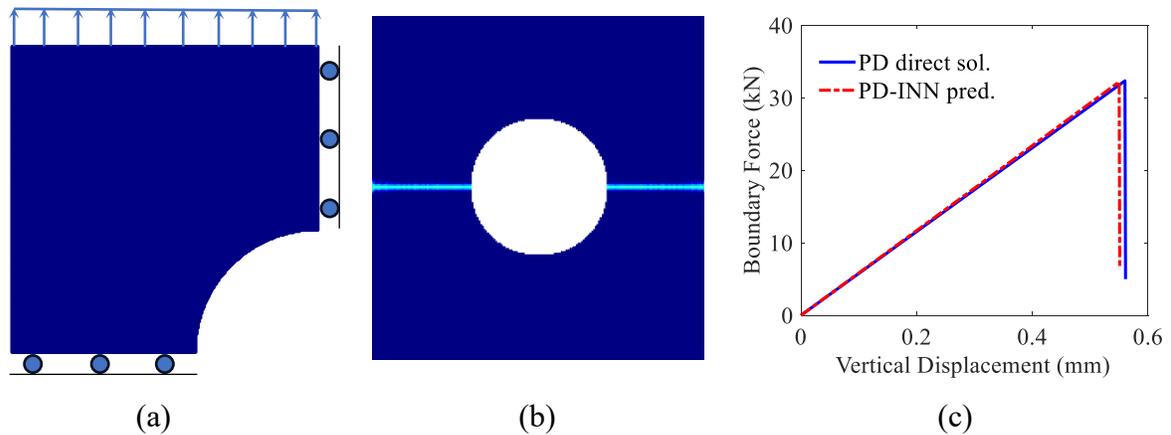

Figure 8. (a) schematic of the model and boundary conditions (b) crack path presentation (c) comparison of the PD-INN and PD numerical solution

In Figure 8-(b), the final crack path is depicted, demonstrating the expected crack propagation originating from regions with high stress concentrations. For visual purposes, the domain is presented with a 90-degree rotational pattern. Figure 8-(c) compares the load-displacement curve obtained from PD-INN with from direct PD numerical solution. The small variations observed between these two results can be attributed to the value of threshold loss considered in this analysis. Despite these variations, our PD-INN demonstrates its capacity to accurately capture the correct crack growth, showcasing a satisfactory level of agreement when compared with the results obtained from direct numerical method.

## 5. Conclusion

This study investigates the potential of combining PD and PINN methods to study crack propagation in brittle structures. By incorporating the governing equations of peridynamics into the neural network's loss function, the proposed PD-INN has the capability to predict deformations in the presence of structural discontinuities. Several case studies involving plates with pre-existing cracks or circular cutout have been examined to evaluate the PD-INN's performance and validate its predictions. It has been shown that the conventional fully connected neural network (FCNN) architecture struggles to interpret geometric features such as pre-cracks within the domain, resulting in convergence failure. Therefore, different PD-INN architectures have been evaluated to discover the network with the potential of crack prediction.

The PD-INN model introduced in this study has maintained the significance of both accuracy and computational efficiency. To have a balance between these two aspects, a novel approach that integrates the residual of the linearized PD equation into the loss function has been introduced. This combination, enhanced by well-adjusted hyperparameters, has led to significant improvements in the effectiveness of the PD-INN. A substantial reduction in computational time has been achieved, by an approximately 12-fold acceleration of convergence rate, showing the practicality and feasibility of this approach. Moreover, a challenge associated with employing PD-INN for a high-density domain, wherein the risks arise converging towards the trivial solution, has been addressed through the implementation of a gradient-aware technique.

To further enhance efficiency in problems associated with damage evolution, the technique of transfer learning has been successfully applied. In addition, a partial training strategy for subsequent time increments has been implemented where only parameters associated with output layer are tuned. These combined advancements not only verify the effectiveness of the proposed approach but also extend the horizons of PD and PINN, promising robust solutions in computational fracture mechanics applications.